\documentclass[prl,twocolumn,superscriptaddress,floatfix]{revtex4-2}

\usepackage{graphicx}
\usepackage{bm}
\usepackage{xcolor}
\usepackage{amsmath}
\usepackage{amssymb}
\usepackage{hyperref}

\begin{document}

\title{Hidden Frustration in Double-Perovskite CaFeTi$_2$O$_6$}

\author{Di Liu}
\thanks{These authors contribute equally to this work.}
\affiliation{Beijing National Laboratory for Condensed Matter Physics, Institute of Physics, Chinese Academy of Sciences, Beijing 100190, China}
\affiliation{School of Physical Sciences, University of Chinese Academy of Sciences, Beijing 100190, China}

\author{Si Wu}
\thanks{These authors contribute equally to this work.}
\affiliation{Institute of Applied Physics and Materials Engineering, University of Macau, Avenida da Universidade, Taipa, Macau, China}

\author{Xuanyu Long}
\thanks{These authors contribute equally to this work.}
\affiliation{Institute for Advanced Study, Tsinghua University, Beijing 100084, China}

\author{Yao Wang}
\affiliation{Beijing National Laboratory for Condensed Matter Physics, Institute of Physics, Chinese Academy of Sciences, Beijing 100190, China}
\affiliation{School of Physical Sciences, University of Chinese Academy of Sciences, Beijing 100190, China}

\author{Hai-Feng Li}
\affiliation{Institute of Applied Physics and Materials Engineering, University of Macau, Avenida da Universidade, Taipa, Macau, China}

\author{Ming Yang}
\affiliation{Wuhan National High Magnetic Field Center and School of Physics, Huazhong University of Science and Technology, Wuhan 430074, China} 

\author{Junfeng Wang}
\affiliation{Wuhan National High Magnetic Field Center and School of Physics, Huazhong University of Science and Technology, Wuhan 430074, China} 

\author{Zheng Liu}
\affiliation{Institute for Advanced Study, Tsinghua University, Beijing 100084, China}
\affiliation{State Key Laboratory of Low-Dimensional Quantum Physics, Department of Physics, Tsinghua University, Beijing 100084, China}

\author{Xiang Li}
\email{xiangli@bit.edu.cn}
\affiliation{Key Laboratory of Advanced Optoelectronic Quantum Architecture and Measurement, Ministry of Education (MOE), School of Physics, Beijing Institute of
Technology, Beijing 100081, China}

\author{Yuan Wan}
\email{yuan.wan@iphy.ac.cn}
\affiliation{Beijing National Laboratory for Condensed Matter Physics, Institute of Physics, Chinese Academy of Sciences, Beijing 100190, China}
\affiliation{School of Physical Sciences, University of Chinese Academy of Sciences, Beijing 100190, China}
\affiliation{Songshan Lake Materials Laboratory, Dongguan, Guangdong 523808, China}

\author{Shiliang Li}
\email{slli@iphy.ac.cn}
\affiliation{Beijing National Laboratory for Condensed Matter Physics, Institute of Physics, Chinese Academy of Sciences, Beijing 100190, China}
\affiliation{School of Physical Sciences, University of Chinese Academy of Sciences, Beijing 100190, China}
\affiliation{Songshan Lake Materials Laboratory, Dongguan, Guangdong 523808, China}

\begin{abstract}
We study the magnetic properties of CaFeTi$_2$O$_6$ (CFTO) by high-field magnetization and specific heat measurements. While the magnetic susceptibility data yield a vanishingly small Curie-Weiss temperature, the magnetic moments are not fully polarized in magnetic field up to 60 T, which reveals a large spin exchange energy scale. Yet, the system shows no long range magnetic order but a spin-glass-like state below 5.5 K in zero field, indicating strong magnetic frustration in this system. Applying magnetic field gradually suppresses the spin-glass-like state and gives rise to a potential quantum spin liquid state whose low-temperature specific heat exhibits a $T^{1.6}$ power-law. Crucially, conventional mechanisms for frustration do not apply to this system as it possesses neither apparent geometrical frustration nor exchange frustration. We suggest that the orbital modulation of exchange interaction is likely the source of hidden frustration in CFTO, and its full characterization may open a new route in the quest for quantum spin liquids.
\end{abstract}

\date{\today}

\maketitle

The road toward quantum spin liquids (QSLs) is paved by the physics of \emph{frustration}~\cite{RamirezAP94,Moessner2006,BalentsL10,SavaryL17,ZhouY17,WenXG19}. In a frustrated magnet, mutually conflicting spin exchange interactions prevent the system from developing long-range magnetic order, thereby opening the window for QSL ground states. A well-established mechanism for frustration is \emph{geometrical} in nature. On one hand, in kagome and pyrochlore lattices~\cite{ShoresMP05,HanTH12,NormanMR16,FengZL17,GaoB19,GaudetJ19}, spin configurations are under-constrained due to the interplay between the Heisenberg exchange interaction and the lattice geometry. The QSLs emerge from the quantum fluctuations among these configurations. On the other hand, although magnets with ideal triangular and square lattices exhibit long range magnetic order, adding further range interactions effectively modify the lattice geometry and may produce QSLs~\cite{LiYS15,ShenY16,PaddisonJAM17,LiuW18,BordelonMM19}. Alternative to the geometrical frustration is the \emph{exchange frustration}. There, strong spin-orbital coupling (SOC) gives rise to significant non-Heisenberg interactions, which are bond-dependent and thus in conflict with each other. A prominent example is the Kitaev magnet, which hosts QSLs despite its bipartite honeycomb lattice structure~\cite{KitaevA06,BanerjeeA17,KasaharaY18,KitagawaK18}.

\begin{figure}
\includegraphics[width=\columnwidth]{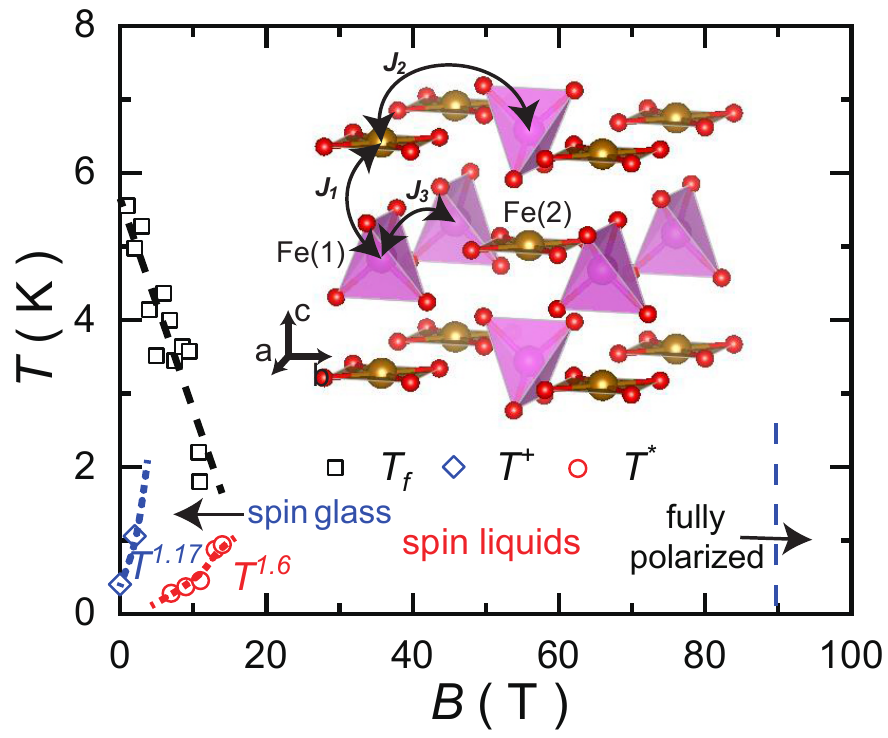}
 \caption{Magnetic phase diagram of CFTO. The dash lines are guides to the eye. The power laws are from the temperature dependence of the specific heat. The inset gives the crystal structure that only shows iron and the associated oxygen ions. The arrows indicate the exchanges between Fe$^{2+}$ ions. }
\label{phasediagram}
\end{figure}

In this work, we put forward a QSL candidate material with neither apparent geometric frustration nor significant exchange frustration  --- CaFeTi$_2$O$_6$ (CFTO). It has the double-perovskite structure with the space group $P4_2/nmc (No. 137)$~\cite{LeinenweberK95a,LeinenweberK95b}. The magnetic ions Fe$^{2+}$ form a tetragonal lattice. Fe$^{2+}$ come in two species that are distinguished by their local environments, namely the tetrahedral Fe(1)O$_4$ and square-planar Fe(2)O$_4$. They are placed alternately along the $c$ axis (Fig. \ref{phasediagram}, inset). Previous magnetic-susceptibility measurement and M\"{o}ssbauer spectroscopy show both species are in the high-spin state with $S$ = 2~\cite{ReiffWM96}. The measured magnetic moment is fairly close to the spin only value, indicating the weak SOC typical for Fe$^{2+}$~\cite{LiX18}. The Curie-Weiss temperature $\theta \approx 0$, which is previously thought to be due to weak exchange and magnetic dipole-dipole interactions~\cite{LiX18}. There is no evidence for magnetic ordering down to 50 mK, but a spin-glass-like transition is found at $T_f \approx$ 5 K~\cite{LiX18}.

The nominally small frustration ratio $\theta/T_f$ seems to suggest that CFTO is far from being a QSL. Surprisingly, we find the spin exchange is actually much larger than $\theta$, by tracing the magnetization curve up to 60 T. Taking the Zeeman energy at the saturation field as an order-of-magnitude measure for the antiferromagnetic spin exchange $J$ yields a frustration ratio $J/T_f\sim O(10)$, thereby putting CFTO in the same category as many familiar geometrically frustrated magnets \cite{RamirezAP94,BalentsL10,SavaryL17}. Furthermore, the spin freezing temperature $T_f$ decreases with increasing magnetic field and is estimated to disappear at $B_c\sim $ 20 T, which points toward a possible field-induced QSL state (Fig.~\ref{phasediagram}). The most interesting aspect of this system is that we find \emph{no} apparent source of frustration that would explain the large frustration ratio, as shown by our detailed theoretical analysis. 

Polycrystalline CFTO samples are synthesized by the high-pressure and high-temperature method as previously reported~\cite{LiX18}. The magnetic susceptibility and specific heat are measured by the magnetic properties measurement system (MPMS, Quantum Desgin) and physical property measurement system (PPMS, Quantum Design) with the dilution-refrigerator insert, respectively. The high-field magnetic moment is measured at Wuhan National High Magnetic Field Center, China. 

\begin{figure}
\includegraphics[width=\columnwidth]{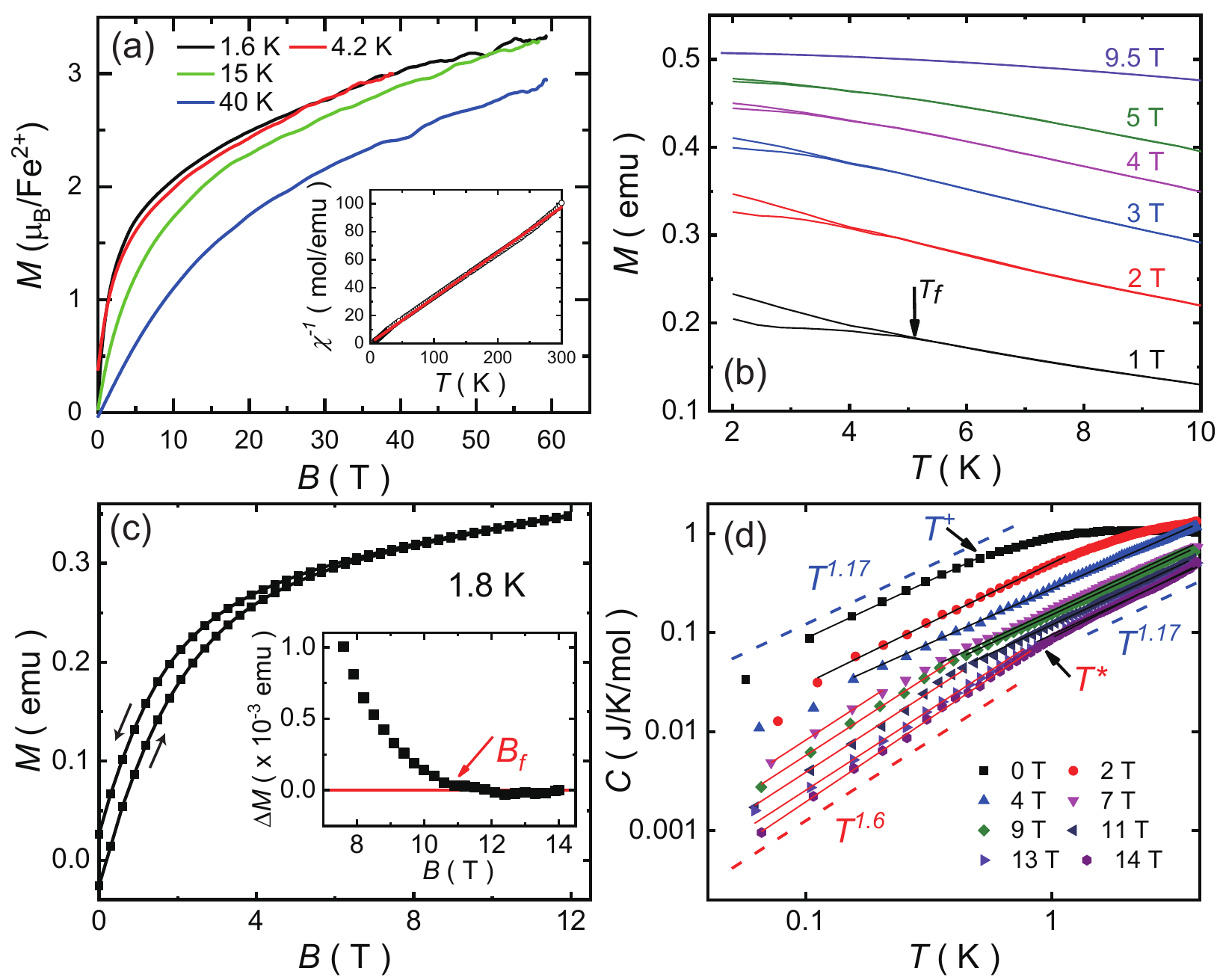}
 \caption{ (a) Field dependence of magnetization at high fields for 1.6, 4.2, 15 and 40 K. The values have been normalized according to low-field data measured at the MPMS. The inset shows the inverse magnetic susceptibility $\chi^{-1}$ = $H/M$ at 1000 Oe as a function of the temperature. The solid line is the fitted result by a linear function. (b) Temperature dependence of magnetization for ZFC and FC processes at different fields. The arrow indicates $T_f$ at 1 T. (c) Field dependence of magnetization at 1.8 K. The arrows indicate the increasing or decreasing field process. The inset shows the field dependence of the difference between the decreasing and increasing field processes. (d) Temperature dependence of the specific heat at different fields. The solid lines are fitted results for the power law function as described in the text. $T^+$ indicates the temperature below which $C \propto T^{1.17}$.  $T^*$ indicates the crossover temperature between the $T^{1.6}$ and $T^{1.17}$ regions.}
\label{experiments}
\end{figure}

The inset of Fig.~\ref{experiments}(a) shows the temperature dependence of the inverse magnetic susceptibility $\chi^{-1}$ of our CFTO sample. The data can be well fitted by the linear function, which corresponds to the Curie-Weiss law $\chi$ = $C/(T-\theta)$. From the Curie constant $C$, we deduce the effective magnetic moment per Fe$^{2+}$ is 4.73 $\mu_B$, which is fairly close to the spin-only value (4.9 $\mu_B$) of an Fe$^{2+}$ in the high spin state ($S=2$). This suggests the weak SOC for Fe$^{2+}$ and high quality of the sample. The Curie-Weiss temperature $\theta\approx1.8$ K, consistent with the previous report~\cite{LiX18}. 

The small Curie-Weiss temperature was previously attributed to the weak spin exchange/magnetic dipole-dipole interactions~\cite{LiX18}. However, one should be alerted that since $\theta$ is proportional to the algebraic sum of exchange constants, for a system containing both ferro- and antiferromagnetic interactions, $\theta$ is not a correct indicator for spin exchange energy scale. To determine the real strength of the exchange interactions, we measured the sample's magnetization $M$ in high magnetic field (Fig.~\ref{experiments}(a)). At 1.6 K, $M$ reaches $\sim 83\%$ of its saturation value ($\sim 4 \mu_B$) at 60 T. This clearly demonstrates that the exchange interactions in this system are \emph{not} weak; otherwise, the spins would have been fully polarized. The small $\theta$ is due to the near compensation of the ferro- and antiferromagnetic exchange interactions. A linear extrapolation of the magnetization curve at 1.6 K yields a crude estimate for the saturation field $B_\mathrm{st} \approx 90$ T. By using the Zeeman energy at $B_\mathrm{st}$, we estimate the order of magnitude of the antiferromagnetic exchange interaction energy scale $J\sim O(10^2) K$, which is two orders of magnitude greater than $\theta$. We also note that the magnetization curve exhibits a relatively rapid increase in weak magnetic field ($\lessapprox 10$ T) resembling a ferromagnet, followed by a slow, quasi-linear growth in stronger fields characteristic of an antiferromagnet.

Figure~\ref{experiments}(b) shows the field-cooled (FC) and zero-field-cooled (ZFC) magnetization as a function of temperature at various fields $B$. At $B\le 5$ T, the FC and ZFC data bifurcate at low temperatures, below which the system enters a glass-like phase~\cite{LiX18}. We define the spin freezing temperature $T_f$ as the onset temperature of the bifurcation. This yields $T_f\approx$ 5 K at 1 T, in agreement with the previous report~\cite{LiX18}. Likewise, the spin freezing also manifests itself in the magnetic hysteresis (Fig.~\ref{experiments}(c)). We define the spin freezing field $B_f$ as the point at which the upward and downward curves bifurcate (Fig.~\ref{experiments}(c), inset). We find $B_f \approx $11 T at 1.8 K.

The large exchange energy scale $J$ and the relatively small value of $T_f$ implies that CFTO is a highly frustrated magnet with a frustration ratio $J/T_f\sim O(10)$. We now study its thermodynamic properties in more detail. Fig.~\ref{experiments}(d) shows the low-temperature specific heat $C$ at various fields. The specific heat shows power-law behavior below $T^+$. The power-law onset temperature $T^+$ increases with the field and eventually becomes larger than 4 K at 4 T, and thus cannot be reliably determined due to the increasing phonon contribution to the specific heat. The power law exponent $\alpha = 1.17$ at zero field. Interestingly, a new exponent $\alpha = 1.6$ emerges at 7 T. The crossover temperature $T^*$ between these two power laws increases with increasing field.

The specific heat of a spin glass is known to exhibit a broad peak above the glass transition temperature, followed by a power-law like behavior as temperature decreases~\cite{BinderK86}. By contrast, the specific heat of CFTO exhibits a peak at $\approx 1.2K$, which is far \emph{below} the spin freezing temperature $T_f$~\cite{LiX18}. Therefore, we think the low-temperature power law behavior observed in the specific heat of CFTO is unlikely due to the spin-glass physics, especially at high fields.

We construct a preliminary temperature-field phase diagram of CFTO based on the above analysis (Fig.~\ref{phasediagram}). The system is in a cooperative paramagnetic state above $T_f$, below which it enters a spin-glass-like state. $T_f$ decreases with increasing field, and, by a linear extrapolation, is expected to vanish at approximately $20$ T. This field could be even smaller as indicated by the emergence of $T^{1.6}$ behavior for the specific heat below $T^*$. Nevertheless, above this field, but below the saturation field $\sim 90$ T, there is a wide window in which the spins are not frozen but also far from being fully polarized. Given that we do not find signatures for magnetic ordering in this window as shown by the magnetization curve in Fig. \ref{experiments}(a), we think the system is potentially a QSL. In particular, we tentatively attribute the power-law behavior of the specific heat with the exponent $\alpha = 1.6$ to the potential proximate QSL that appears at higher field. 

\begin{figure}
\includegraphics[width=\columnwidth]{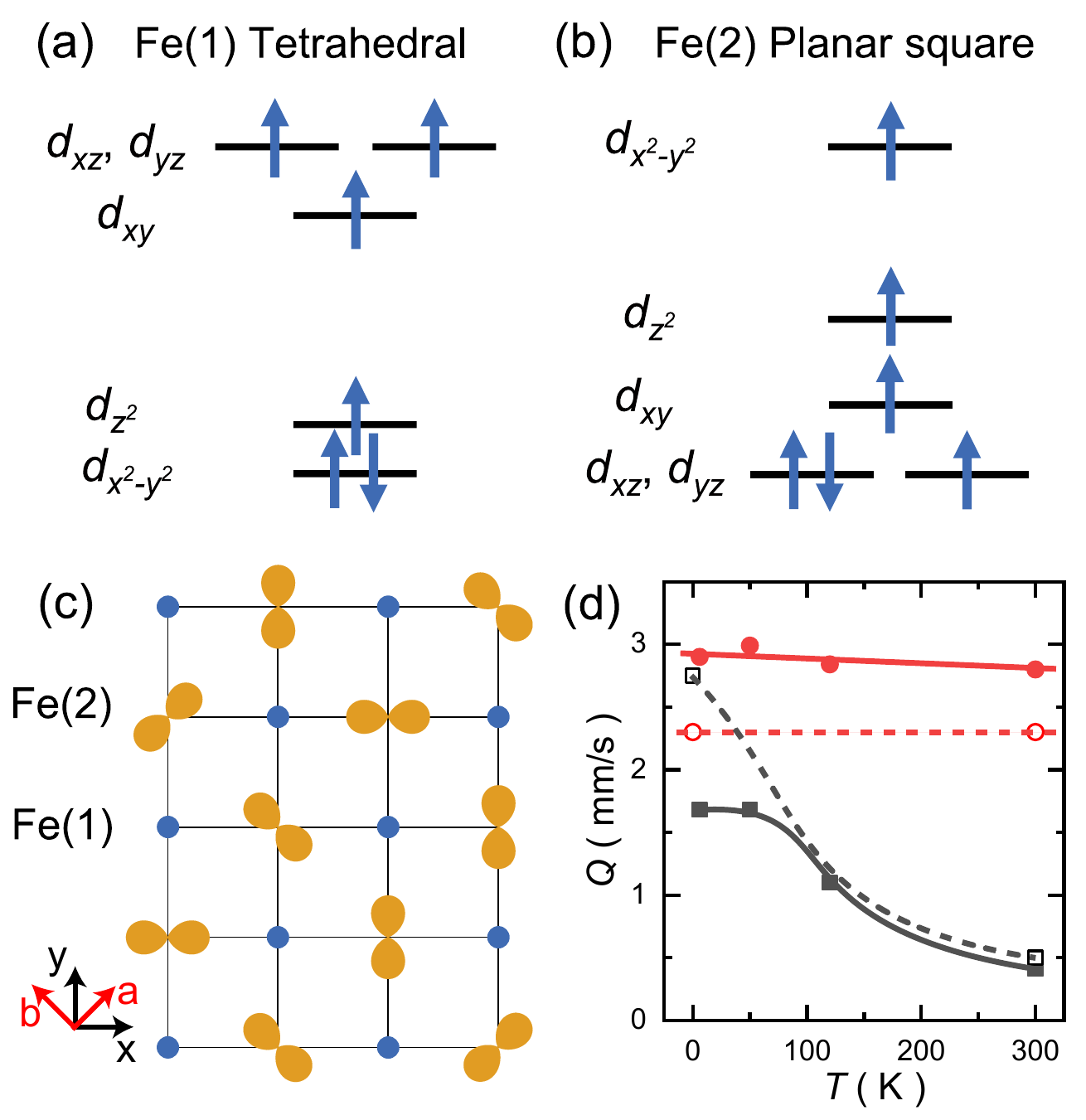}
 \caption{(a), (b) The crystal field splitting and electron filling of the two types of Fe$^{2+}$ in CFTO. We note that the natural $x$, $y$ axes of the $d$-orbitals compatible with the crystal fields are rotated by 45 degrees from the crystallographic $a$, $b$ axes. (c) Schematic of the four types of local orbital configurations at Fe(2) sites converged by the DFT calculations~\cite{note1}. Note that the $x-y$ axes in theoretical calculations are rotated 45 degrees relative to the lattice axes as shown by the arrows. The dumbbells represent the charge density of the occupied orbital. If the orientations of the dumbbells are parallel to the $x$ or $y$ axis, the strength of exchange interactions between Fe(1) and Fe(2) along $x$ or $y$ direction becomes different. (d) Temperature dependence of QS for Fe(1) (red circles) and Fe(2) (black squares). The filled and open symbols represent experimental \cite{LiX18} and theoretical results~\cite{note1}, respectively. Here theoretical values with and without orbital polarizations are put at 0 and 300 K, respectively. The lines are guides to the eye.}
\label{DFT}
\end{figure}

Viewing from the microscopic angle, a main mystery of CFTO is its frustration mechanism. A few familiar scenarios that may lead to large frustration ratio can either be excluded or are unlikely to occur in this system:

Firstly, it is known that dimensional reduction may lead to a much lower magnetic ordering temperature. Here, although the relative short Fe-Fe distance along the $c$ axis [labeled as $J_1$ in the inset of Fig.~\ref{phasediagram}] renders the possibility of weakly coupled spin chains, it would be difficult to reconcile the spin chain picture with the small Curie-Weiss temperature $\theta$. 
    
Secondly, the fairly weak SOC in this system precludes the exchange frustration scenario. Furthermore, while the subtle interplay between the SOC and the $e_g$ orbital manifold may produce a unique spin-orbital entangled singlet state and thereby result in spin-liquid-like behavior in some Fe$^{2+}$-based systems \cite{FritschV04,KrimmelA05,GangC09,PlumbKW16,SettyC17},  it is unlikely to occur in CFTO due to distinct crystal field environments and orbital characters (see below for details).  

Thirdly, there is no apparent geometrical frustration in this system. As we have shown earlier, the small Curie-Weiss temperature is due to compensation. This necessarily entails two kinds of spin exchange interactions. Based on the crystal structure, the natural choices would be the exchange interactions between nearest-neighbors ($J_1$) and second-neighbors ($J_2$) [Fig.~\ref{phasediagram}(a)].  With only $J_1$ and $J_2$, the spins form a bipartite tetragonal lattice, which is geometrically unfrustrated. Frustration due to longer range exchange interactions, e.g. competition between $J_2$ and $J_3$ within the $ab$-plane, is unlikely since, in a good insulator such as CFTO, the exchange strength is expected to decay exponentially as Fe-Fe distance increases. Indeed, preliminary DFT calculations suggest $J_1$ and $J_2$ are the dominant exchange interactions in this system \cite{note1}. 

An important observation is that, the isostructural CaMnTi$_2$O$_6$ (CMTO), where Mn$^{2+}$ replaces Fe$^{2+}$, is clearly free from frustration. Its Curie-Weiss temperature $\theta_{CW}=-32 K$, and it shows magnetic ordering at $T_N = 10$ K \cite{AimiA14}. A key difference between these two siblings lies in the fact Fe$^{2+}$ possesses active orbital degrees of freedom whereas Mn$^{2+}$ does not.

To see this, we revisit the elementary electronic structure. Due to the distinct oxygen configurations (Fig. \ref{phasediagram}, inset), Fe(1) and Fe(2) experience different crystal field splittings. The $d$ orbitals of the tetrahedral Fe(1) first split into a higher $t_\mathrm{2g}$ triplet and a lower $e_\mathrm{g}$ doublet, and the lower $e_\mathrm{g}$ doublet then split into d$_{z^2}$ and d$_{x^2-y^2}$ singlets by further neighbor O$^{2-}$. The $d$ orbitals of the planar-square Fe(2) split into three higher singlets and a lower doublet consisting of d$_{xz}$ and d$_{yz}$ orbitals. The degeneracy between the d$_{xz}$ and d$_{yz}$ orbitals is protected by the tetragonal crystal symmetry. The lowest-energy electron configurations according to the Hund's rules are plotted in Fig.~\ref{DFT}(a) and Fig.~\ref{DFT}(b), for Fe(1) and Fe(2), respectively. Note that for Fe(2), since the lowest levels are doubly degenerate, the orbital assignment to the spin-down electron is not unique, thereby introducing an extra orbital degree of freedom. By contrast, in CMTO, this orbital degree of freedom does not exist because Mn$^{2+}$ has only five electrons.

Breaking the tetragonal symmetry of the lattice lifts the degeneracy between the $d_{xz}/d_{yz}$ orbitals. Indeed, by allowing a \emph{spontaneous} breaking of the said symmetry on Fe(2), the DFT calculations can successfully reproduce the insulating state of the CFTO~\cite{note1}. We are able to target insulating solutions with various orbital ordering patterns [Fig.~\ref{DFT}(c)], all of which are lower in energy than the previously reported half-metallic state by $O(10)$ meV per Fe~\cite{15DFT}. The solutions either select $d_{xz}$ or $d_{yz}$ state out of the degenerate doublet, or the linear combination $(d_{xz}\pm d_{yz})/\sqrt{2}$.

The spontaneous breaking of the $d_{xz}/d_{yz}$ doublet modifies the local electrostatic environment, which can be measured by M\"{o}ssbauer spectroscopy through the electric quadrupole splitting (QS). Without orbital polarization, DFT calculations ~\cite{note1} yield two QS values that are consistent with the experimental ones at room temperature (Fig. \ref{DFT}(d)). In particular, we identify the smaller QS being due to Fe(2) and the larger QS to Fe(1). We note that previous experimental study assigned the smaller QS to the Fe(1) site~\cite{LiX18}, but this assignment is not supported by DFT. With orbital polarization, DFT calculations show that the QS for Fe(2) increases drastically, which explains the experimental observation that the smaller QS value increases with temperature. By contrast, Fe(1) has an almost fixed QS. As no long-range orbital ordering or the associated structural transition has been detected so far, we expect that the orbitals are either quenched to a random distribution or fluctuating in a correlated liquid-like state as the temperature decreases. 

We thus turn to the orbital degrees of freedom on Fe(2) to seek a possible frustration mechanism. The orbital polarization naturally modulates $J_2$, which can be intuitively understood in terms of wave function overlap --- If the electrons on a Fe(2) site are in the $d_{xz}$ ($d_{yz}$) state, their overlap with the neighboring Fe(1) electrons is larger along the $x$ ($y$) axis [Fig.~\ref{DFT}(c)]. The local structural distortion that accompanies with the orbital splitting will further enhance the modulation. 

This orbital modulation effect can be captured phenomenologically by the following interaction:
\begin{align}
H' = g J_2 \sum_i \tau_i \mathbf{S}_i \cdot (\mathbf{S}_{i+\hat{x}}+\mathbf{S}_{i-\hat{x}}-\mathbf{S}_{i+\hat{y}}-\mathbf{S}_{i-\hat{y}}).
\label{eq:modulation}
\end{align}
$\tau_i$ describes the orbital degrees of freedom on Fe(2) site $i$: $\tau_i  = 1$ ($\tau_i = -1$) if the electrons occupy the $d_{xz}$ ($d_{yz}$) orbital; $\tau_i = 0$ if the electrons occupy the $(d_{xz}\pm d_{yz})/\sqrt{2}$ orbital.  The dimensionless parameter $g$ describes the orbital modulation of $J_2$ exchange interaction. A crucial consequence is that the orbital fluctuations can switch the sign of the exchange interactions in the $xy$ plane when $|g|>1$, thereby giving rise to frustration. Analysis of the classical spin model with $J_1,J_2$ exchange, and $H'$ shows that the maximal frustration effect is reached at $|g| = 1$, at which point the system possesses a large family of sub-extensively degenerate ground states \cite{note1}. 

We note that SOC has been proposed to play a crucial role in the possible QSL in FeSc$_2$S$_4$ through the competition between on-site spin-orbit coupling and Kugel-Khomskii exchange \cite{FritschV04,KrimmelA05,GangC09}. Although later studies show that the system is magnetically ordered and structurally distorted at low temperatures, the essence of SOC physics remains \cite{PlumbKW16,SettyC17}. Here for CaFeTi$_2$O$_6$, the situation is very different because the orbital moment is most likely quenched as suggested by the effective moment. More importantly, there are two Fe sites with completely different local symmetries, which makes both the orbitals and superexchanges complicated (see below for detailed discussions). It is extremely hard to find a tricky balance between these complex interactions to give a non-ordered state.

In summary, we have argued that CFTO is a highly frustrated magnet that potentially hosts a field-induced QSL. Crucially, its frustration mechanism does not fit into the framework of either geometrical or exchange frustration. We hypothesis that the magnetic frustration observed in CFTO has an orbital origin as described by Eq.~\eqref{eq:modulation}. This mechanism is reminiscent of the spin-lattice liquid state in Y$_2$Mo$_2$O$_7$, which is thought to be induced by subtle local structural distortions in conjugation with large magnetoelastic effect~\cite{Smerald2019}. Looking ahead, it will be important to search for direct evidence for the orbital disorder/fluctuation and to quantify the orbital modulation effect to fully unveil this hidden frustration.

\begin{acknowledgments}

We thank George Jackeli for helpful discussions. This work is supported by the National Key Research and Development Program of China (Grants No.~2017YFA0302900 and No.~2016YFA0300502), the National Natural Science Foundation of China (Grants No.~11961160699, No.~11874401, No.~11974396, and No.~11774196), the K. C. Wong Education Foundation (Grant No.~GJTD-2020-01), the Strategic Priority Research Program of the Chinese Academy of Sciences (Grant No.~XDB33000000), and Tsinghua University Initiative Scientific Research Program.

\end{acknowledgments}

\end{document}